\documentclass{mn2e}
\usepackage{graphicx}
\begin{document}

\title[Disentangling jet and disc emission from the 2005 outburst of XTE J1118+480]
      {Disentangling jet and disc emission from the 2005 outburst of XTE J1118+480}
\author[Brocksopp et al.]
    {C.~Brocksopp$^1$\thanks{email: cb4@mssl.ucl.ac.uk}, 
 P.G.~ Jonker$^{2,3}$, D.~Maitra$^{4,5}$, H.A.~Krimm$^{6,7}$, G.G.~Pooley$^8$, G.~Ramsay$^9$,
\newauthor
C.~Zurita$^{10}$\\
$^1$Mullard Space Science Laboratory, University College London, Holmbury St. Mary, Dorking, Surrey RH5 6NT\\
$^2$SRON, Netherlands Institute for Space Research, 3584 CA Utrecht, The Netherlands\\
$^3$Harvard--Smithsonian Center for Astrophysics, 60 Garden Street, Cambridge, MA~02138, U.S.A.\\
$^4$Astronomical Institute ``Anton Pannekoek'' and Center for high-Energy Astrophysics, University of Amsterdam, Kruislaan 403, \\
Amsterdam 1098 SJ, The Netherlands\\
$^5$Department of Astronomy, University of Michigan, Ann Arbor, Michigan 48109, USA\\
$^6$NASA/Goddard Space Flight Centre, Greenbelt, MD 20771, USA\\
$^7$Universities Space Research Association, Columbia, MD 20144, USA\\
$^8$Mullard Radio Astronomy Observatory, Cavendish Laboratory, Madingley Road, Cambridge CB3 0HE\\
$^9$Armagh Observatory, College Hill, Armagh BT61 9DG\\
$^{10}$Instituto de Astrofisica de Canaris, 38200 LaLaguna, Tenerife, Spain\\
}
\date{Accepted ??. Received ??}
\pagerange{\pageref{firstpage}--\pageref{lastpage}}
\pubyear{??}
\maketitle
\begin{abstract}

The black hole X-ray transient, XTE~J1118+480, has now twice been observed in outburst --- 2000 and 2005 --- and on both occasions remained in the low/hard X-ray spectral state. Here we present radio, infrared, optical, soft X-ray and hard X-ray observations of the more recent outburst. We find that the lightcurves have very different morphologies compared with the 2000 event and the optical decay is delayed relative to the X-ray/radio. We attribute this lesser degree of correlation to contributions of emission from multiple components, in particular the jet and accretion disc. Whereas the jet seemed to dominate the broadband spectrum in 2000, in 2005 the accretion disc seems to be more prominent and we use an analysis of the lightcurves and spectra to distinguish between the jet and disc emission. There also appears to be an optically thin component to the radio emission in the 2005 data, possibly associated with multiple ejection events and decaying as the outburst proceeds. These results add to the discussion that the term ``low/hard state'' covers a wider range of properties than previously thought, if it is to account for XTE~J1118+480 during these two outbursts. 
\end{abstract}

\section{Introduction}
XTE~J1118+480 is a black hole X-ray transient, notable for being one of a small subset of transient systems observed in the low/hard X-ray spectral state throughout at least one outburst (e.g. Brocksopp et al. 2004). The low/hard state is defined in terms of the X-ray spectrum; traditionally the X-ray emission at 2--20 keV could be well-described by a power-law, with the photon index, $\Gamma$, in the range $1.5<\Gamma <2.1$ and with no or little need for a disc component (McClintock \& Remillard 2006). This is consistent with the X-ray emission in this energy range being produced via Compton up-scattering of soft photons by a hot electron plasma (e.g. corona, jet or advection-dominated accretion flow) and, indeed, a partially self-absorbed radio jet has now long been associated with the low/hard state (e.g. Fender 2006 and references therein). 

Controversially, XTE ~J1118+480 then became the first source for which a compact jet model could be fit to the {\em whole} spectrum, from radio to hard X-rays, and required a disc contribution only in the optical and ultraviolet regions (Markoff, Falcke, Fender 2001; see also, however, alternative explanations in Zdziarski et al. 2003, Zdziarski \& Gierlinski 2004 and revised jet models in e.g. Markoff, Nowak \& Wilms 2005, Maitra et al. 2009). The implications of a dominant X-ray synchrotron and/or inverse Compton emission component for the energy budget of the jet are far-reaching and, if confirmed, would necessitate significant changes to current models for accretion discs and their outbursts.

Similar jet models have been used to describe low/hard state spectra of GX~339$-$4 and Cyg X-1 (Markoff, Nowak, Wilms 2005). However, GX~339$-$4, Cyg X-1 and the 2000 outburst of XTE~J1118+480 have had unusually long periods when a stable jet is present; the presence of jets is more commonly restricted to brief phases of a transient outburst (Fender, Belloni, Gallo 2004). In such cases the canonical self-absorbed radio spectrum of the low/hard state is likely to be ``contaminated'' with optically thin radio emission (e.g. Jonker et al. 2009). While we have a reasonably good understanding of the sequence of progression from one spectral state to another, evolution of the jet over the course of a low/hard state outburst has not yet been studied in detail. We do not yet know whether a jet, emitting from radio to X-ray wavelengths, is a typical feature of transient outbursts or a rare phenomenon, occuring only under extreme conditions in some specific sources or events.

\begin{table}
\begin{center}
\caption{Hard and soft X-ray count rates from {\sl Swift} (top) and {\sl RXTE} (bottom). The energy ranges for the BAT, PCA and HEXTE are 14--195, 3--20 and 20--200 keV respectively.}
\label{tab:xcounts}
\begin{tabular}{lc}
\hline
\hline
MJD &BAT (cts/s/cm$^2$) \\
\hline
53380 &      $0.0024      \pm     0.0006$      \\         
53382 &      $0.0073      \pm     0.0003$      \\         
53384 &      $0.0102      \pm     0.0003$       \\        
53388 &      $0.0067      \pm     0.0007$      \\         
53390 &      $0.0058      \pm     0.0003$      \\         
53392 &      $0.0057      \pm     0.0006$         \\      
53394 &      $0.0033      \pm     0.0007$      \\         
53396 &      $0.0020      \pm     0.0007$        \\       
53398 &      $0.0020      \pm     0.0004$       \\        
53404 &      $0.0012      \pm     0.0007$       \\   
\hline
\end{tabular}
\begin{tabular}{lcc}
\hline
MJD & PCA (cts/s)&HEXTE (cte/s)\\
\hline
53383.3 &$56.12\pm 0.36 $& $5.42 \pm 0.66   $      \\      
53384.2 &$57.07\pm 0.30 $& $5.56 \pm 0.62   $     \\       
53385.0 &$55.87\pm 0.17 $& $6.65 \pm 0.21   $     \\       
53385.7 &$53.80\pm 0.24 $& $6.02 \pm 0.34   $      \\      
53385.8 &$51.14\pm 0.16 $& $6.00 \pm 0.23   $       \\     
53386.4 &$47.92\pm 0.22 $& $5.59 \pm 0.31   $        \\    
53386.8 &$44.68\pm 0.15 $& $4.97 \pm 0.23   $        \\    
53387.2 &$45.22\pm 0.21 $& $5.05 \pm 0.33   $        \\    
53387.8 &$42.32\pm 0.14 $& $5.15 \pm 0.22   $        \\    
53388.6 &$42.79\pm 0.16 $& $5.06 \pm 0.23   $         \\   
53388.8 &$39.53\pm 0.14 $& $5.18 \pm 0.21   $         \\   
53389.1 &$37.79\pm 0.28 $& $4.36 \pm 0.84   $         \\   
53389.6 &$38.48\pm 0.19 $& $3.87 \pm 0.32   $         \\   
53390.1 &$36.99\pm 0.26 $& $3.69 \pm 0.43   $         \\   
53390.5 &$35.41\pm 0.20 $& $3.68 \pm 0.34   $         \\   
53391.1 &$31.89\pm 0.20 $& $4.19 \pm 0.37   $          \\   
53392.0 &$28.55\pm 0.23 $& $2.75 \pm 0.39   $        \\    
53392.1 &$29.32\pm 0.23 $& $2.85 \pm 0.41   $     \\       
53392.7 &$24.54\pm 0.24 $& $2.46 \pm 0.43   $     \\                
53393.7 &$20.78\pm 0.09 $& $2.41 \pm 0.17   $     \\                
53394.2 &$19.07\pm 0.16 $& $1.84 \pm 0.35   $     \\                
53394.2 &$18.71\pm 0.15 $& $1.05 \pm 0.31   $     \\                
53394.7 &$16.47\pm 0.12 $& $1.74 \pm 0.25   $     \\                
53395.1 &$15.29\pm 0.17 $& $2.23 \pm 0.42   $     \\                
53395.6 &$14.11\pm 0.12 $& $1.25 \pm 0.27   $     \\                
53396.1 &$12.32\pm 0.16 $& $1.42 \pm 0.39   $     \\                
53398.0 &$8.15\pm 0.17 $ & $  0.99\pm 0.61 $     \\                  
53398.1 &$8.35\pm 0.09 $ & $  0.57\pm 0.28 $     \\                  
53398.4 &$7.95\pm 0.08 $ & $  0.85\pm 0.28 $     \\                  
53399.3 &$6.33\pm 0.16 $ & $  0.41\pm 0.42 $     \\                  
53400.4 &$4.64\pm 0.17 $ & $  0.80\pm 0.45 $     \\                  
53400.5 &$4.54\pm 0.11 $ & $  0.54\pm 0.29 $     \\                  
53400.9 &$4.07\pm 0.14 $ & $  0.04\pm 0.39 $     \\                  
53403.4 &$1.90\pm 0.15 $ & $  0.13\pm 0.41 $     \\      
\hline                               
\hline
\end{tabular}
\end{center}
\end{table}

\begin{table}
\begin{center}
\caption{LT photometry in $B$ and $V$ bands and UKIRT photometry in $J$, $H$ and $K$ bands. Values are given magnitudes.}
\label{tab:optir}
\begin{tabular}{lcc}
\hline
\hline
MJD    &$B$        &$V$        \\
\hline
53393.0 &13.64 $\pm$0.06 & --\\
53393.1 &13.60 $\pm$0.04 & 13.54 $\pm$0.04        \\
53395.0 &13.86 $\pm$0.06 & 13.61 $\pm$0.05 \\
53395.1 &13.90 $\pm$0.05 & 13.60 $\pm$0.04 \\
53396.0 &13.98 $\pm$0.06 & 13.73 $\pm$0.05 \\
53396.1 &13.91 $\pm$0.05 & 13.72 $\pm$0.05  \\
\hline
\end{tabular}
\begin{tabular}{lccc} 
\hline
MJD    &$J$        &$H$        &$K$      \\
\hline
53390.4& 13.08$\pm$ 0.01& 12.66$\pm$ 0.02& 12.17$\pm$ 0.02\\
53390.6& 13.05$\pm$ 0.01& 12.67$\pm$ 0.02& 12.11$\pm$ 0.02\\
53391.4& 13.05$\pm$ 0.01& 12.72$\pm$ 0.02& 12.28$\pm$ 0.02\\
53391.6& 13.14$\pm$ 0.01& 12.80$\pm$ 0.02& 12.24$\pm$ 0.02\\
53393.4& 13.28$\pm$ 0.01& 12.96$\pm$ 0.02& 12.40$\pm$ 0.02\\
53393.7& 13.20$\pm$ 0.01& 12.91$\pm$ 0.02& 12.51$\pm$ 0.02\\
53394.4& 13.29$\pm$ 0.01& 12.95$\pm$ 0.02& 12.55$\pm$ 0.02\\
53394.7& 13.27$\pm$ 0.01& 12.97$\pm$ 0.02& 12.54$\pm$ 0.02\\
53395.4& 13.39$\pm$ 0.01& 13.04$\pm$ 0.02& 12.60$\pm$ 0.02\\
53395.6& 13.34$\pm$ 0.01& 13.04$\pm$ 0.02& 12.65$\pm$ 0.02\\
53409.6& 14.80$\pm$ 0.00& 14.56$\pm$ 0.01& 14.11$\pm$ 0.01\\
53412.5& 15.93$\pm$ 0.01& 15.40$\pm$ 0.01& 15.00$\pm$ 0.01\\
53426.5& 17.06$\pm$ 0.03& 16.24$\pm$ 0.03& 15.97$\pm$ 0.04\\
\hline
\end{tabular}
\end{center}
\end{table}

The relative importance of jet, disc and corona is one of the key questions currently facing the field of X-ray binaries. Markoff et al. (2005) show that high signal-to-noise spectra of Cyg X-1 and GX~339$-$4 can be {\em equally} well fit by jet and Compton models. Hynes et al. (2006) emphasised the difficulties in extrapolating spectral models from the infrared/optical/ultraviolet region to the X-rays, the former of which is typically a mixture of jet and disc emission. The spectral break between flat-spectrum and optically thin emission is potentially hidden by the disc component and so cannot necessarily be observed directly (although see also Russell et al. 2006, Migliari 2008). Instead Hynes et al. (2006) showed that the short-term variability can help to distinguish the components. Alternatively Brocksopp et al. (2006) showed that the long-term variability and (anti-)correlation of multiwavelength datasets can also be crucial to disentangling the contributions from disc, jet and/or corona and we adopt a similar approach here. 

\subsection{XTE J1118+480}
XTE~J1118+480 was discovered in 2000 when it entered its first known outburst (Remillard et al. 2000). The event lasted $\sim7$ months, decaying after a first peak and then rebrightening to a ``plateau state'' for the latter 5 months, staying in the low/hard state throughout (Chaty et al. 2003; Brocksopp, Bandyopadhyay \& Fender 2004). While a synchrotron spectrum at hard X-rays remains controversial, it was independently considered likely in the infrared, optical and ultraviolet regions (Kanbach et al. 2001; Hynes et al. 2003). Frontera et al. (2003) showed that the X-ray spectra could be described by a model of thermal Comptonization plus blackbody, but that the temporal properties were indicative of a non-thermal, and likely synchrotron, component to the soft X-rays. More recently Reis, Miller \& Fabian (2009) discovered a thermal disc component with temperature 0.21 keV in {\sl Chandra} spectra (although see also Gierli\'nski, Done \& Page 2008, 2009). We note that all these observations took place during the second, plateau-like, phase of the outburst with the earlier peak only being discovered retrospectively in {\sl RXTE}/ASM and {\sl CGRO}/BATSE data (Remillard et al. 2000; Wilson \& McCollough 2000).

The second known outburst of XTE~J1118+480 began in 2005 January, discovered at optical wavelengths (Zurita et al. 2005) and confirmed by X-ray and radio observations (Remillard et al. 2005; Pooley 2005). {\sl RXTE} observations confirmed that the X-ray source again remained in the low/hard state (Swank \& Markwardt 2005; Zurita et al. 2006). A further reflare occurred in the optical, superimposed on the decay profile, (Chou et al. 2005) but not the radio (Rupen, Dhawan \& Mioduszewski 2005). Hynes et al. (2006) studied the short-term variability of the infrared, optical and X-ray emission and discovered an optically thin synchrotron origin to the variable infrared component. A similar result was discovered from comparison of the optical and infrared SEDs (Zurita et al. 2006).

\section{Observations}
Observations took place at hard and soft X-rays, optical, infrared and radio with a view to obtaining full spectral coverage for a large number of epochs, as required for monitoring of the spectrum as the outburst evolved.

\begin{table*}
\begin{center}
\caption{VLA flux densities at 1.42, 4.7, 8.4, 15 and 23 GHz. $1\sigma$ errors are listed for detections; $3\sigma$ upper limits are included for non-detections. Ryle Telescope daily-averaged flux densities at 15 GHz are also listed.}
\label{tab:radio}
\begin{tabular}{lccccc}
\hline
\hline
MJD &  \multicolumn{5}{c}{Radio Flux Density (mJy)} \\
             & 1.42~GHz    & 4.7~GHz               & 8.4~GHz                 & 15~GHz                 & 23~GHz \\
\hline
53383.2 &--&--&--& 8.7$\pm$ 1.9&\\ 
53383.3 &     --           &5.85$\pm$ 0.13&12.29$\pm$ 0.15&8.74$\pm$0.70 &7.44$\pm$0.49 \\
53384.0 &--&--&--&11.2$\pm$2.0&\\ 
53384.7 &        --        &5.70$\pm$ 0.20&7.69$\pm$ 0.15 &   --             &      --          \\
53385.4 &--&--&--& 9.6$\pm$ 2.8&\\  
53386.0 &--&--&--& 8.0$\pm$ 2.1&\\ 
53386.2 &2.45$\pm$0.30 &4.44$\pm$ 0.19&6.12$\pm$ 0.13 &       --         &       --         \\
53387.4 &--&--&--& 7.8$\pm$ 2.3&\\ 
53388.4 &--&--&--& 7.8$\pm$ 1.5&\\ 
53391.6 &1.58$\pm$0.27 &3.97$\pm$ 0.17&5.29$\pm$ 0.10 &          --      &5.34$\pm$ 0.32 \\
53393.0 &--&--&--& 3.7$\pm$ 1.9&\\ 
53393.9 &--&--&--& 4.2$\pm$ 1.9&\\ 
53394.3 &          --      &2.74$\pm$ 0.18&3.35$\pm$ 0.11 &      --          &5.18$\pm$ 0.30 \\
53396.1 &--&--&--& 3.4$\pm$ 2.1&\\
 53396.5 &0.76$\pm$0.20 &2.38$\pm$ 0.17&2.57$\pm$ 0.10 &2.69$\pm$0.53 &3.61$\pm$ 0.28 \\
53399.3 &0.50$\pm$0.22 &1.86$\pm$ 0.16&1.81$\pm$ 0.10 &2.05$\pm$0.42 &2.55$\pm$ 0.19 \\
53401.3 &   --             &     --	     &1.55$\pm$ 0.12 &      --          &      --        \\
53404.2 &    --            &0.40$\pm$ 0.13&0.96$\pm$ 0.14 &      --          &1.25$\pm$ 0.24 \\
53407.5 &     --           &0.21$\pm$ 0.10&0.40$\pm$ 0.13 &         --       &        --         \\
53414.0 &    --            &$\le0.13$              &$\le0.13$                &    --            &   --       \\
53421.0 &    --            &$\le0.15$              &$\le0.10$                 &  --             &   --   \\
53428.0 &    --            & ---                       &$\le0.12$                &   --           &  --\\
\hline
\end{tabular}
\end{center}
\end{table*}

\begin{figure*}
\vspace*{-0.5cm}
\begin{center}
\leavevmode
\includegraphics[width=16.5cm]{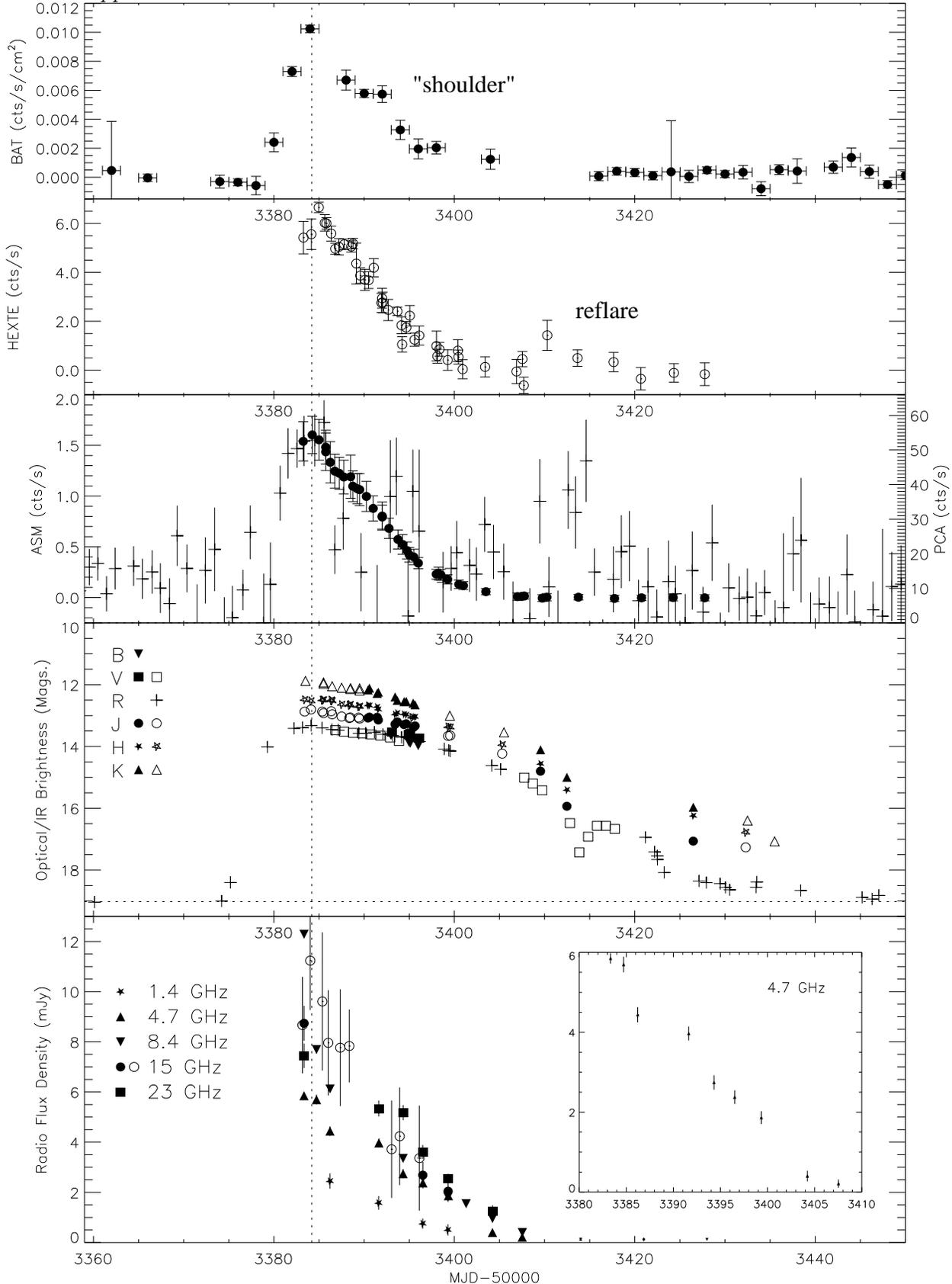}
\vspace*{-0.25cm}
\caption{Lightcurves for each waveband. Top: {\sl Swift}/BAT hard X-rays (14--195 keV). Second: {\sl RXTE}/HEXTE (20--100 keV). Third: {\sl RXTE}/ASM (2--12 keV; crosses) and {\sl RXTE}/PCA (3--20 keV; solid symbols). Fourth: optical and infrared points, from this work (solid symbols) and the literature (other symbols). Horizontal line indicates $R$-band quiescence (Zurita et al. 2006). Error-bars are to within the size of the points. Bottom: radio data from the VLA (solid symbols) and Ryle Telescope (open symbols). Small symbols (after MJD 53410) are $3\sigma$ upper limits at 4.7 and 8.4 GHz. Inset plot shows more clearly the deviation from a smooth decay at 4.7 GHz. The dashed vertical line in all plots represents the time of the PCA peak.}
\label{fig:lightcurves}
\end{center}
\end{figure*}

\subsection{X-rays}
Hard X-ray observations in the energy range 14--195 keV were obtained by the Burst Alert Telescope (BAT) on-board {\sl Swift}. These observations were particularly fortuitous since {\sl Swift} had been launched less than two months prior to the outburst of XTE~J1118+480. The BAT had only been collecting data for a month, making this outburst one of the first transient events observed by {\sl Swift}. The other {\sl Swift} instruments were not yet in full operational mode and no pointed observations were achievable. However, the large field of view of the BAT meant that XTE~J1118+480 was detected on ten occasions while still bright. The spectra were processed in a manner similar to that for GRO~J1655$-$40 (see Brocksopp et al. 2006 for full details) and the count rates listed in Table~\ref{tab:xcounts}.

Further hard X-ray observations in the range 20--200 keV were obtained using the public archive of the HEXTE (High Energy X-ray Timing Experiment) instrument on-board the Rossi X-ray Timing Explorer ({\sl RXTE}). Data from both clusters were added and then binned so as to achieve a minimum of 200 counts/bin using {\sc grppha}. We note that, while detector 2 on Cluster 1 lost its spectral capabilities during the early stages of the mission, this is accounted for during the standard data reduction. Furthermore we have checked the spectra of the plus and minus background positions (for more details see e.g. Section 3 of Rothschild \& Lingenfelter 2003) for our observations and confirmed that there was no noticeable difference between the clusters.

Soft X-ray observations were obtained from the public archives of both the All Sky Monitor (ASM; 2--12 keV) and the Proportional Counting Array (PCA; 2--20 keV) on-board {\sl RXTE}, the latter of which gives better sensitivity but less frequent observations. We extracted spectra from the PCA data using {\sc ftools} version 6.1.1, including data only from the Proportional Counter Unit 2 which was functioning throughout. We used the background model appropriate for faint sources and corrected the spectra for dead-time (although the source count-rate is such that dead-time effects are small). We added a systematic error of 0.6\% to the count-rate in each spectral bin using {\sc grppha}. The resultant PCA count rates are listed in Table~\ref{tab:xcounts}.

\subsection{Optical / Infrared}
$B$- and $V$-band observations were obtained via a Target of Opportunity proposal at the Liverpool Telescope (LT). The RATCam instrument was used and the data were reduced using the standard LT pipeline routines\footnote{http://telescope.livjm.ac.uk/Info/TelInst/Pipelines/}. The resultant apparent magnitudes were extracted using {\sc Gaia}, with reference to observations of standard stars, and are listed in Table~\ref{tab:optir}. Additional optical photometry has been retrieved from the literature (Zurita et al. 2005a,b,c,2006; Chou et al. 2005a,b).

$J$-, $H$- and $K$-band observations were obtained via a Target of Opportunity proposal at the UK Infrared Telescope (UKIRT). The UFTI (UKIRT Fast Track Imager) instrument was used and the data were reduced using the standard UKIRT software, {\sc oracdr}, again with reference to standard stars. The resultant apparent magnitudes are given in Table~\ref{tab:optir}. Additional near-infrared photometry has been extracted from the literature (Hynes et al. 2005, 2006)

\subsection{Radio}
We extracted VLA radio observations at 1.42, 4.7, 8.4, 15 and 23 GHz from the public archive of the Very Large Array (VLA). The array was in the A configuration at the beginning of the outburst, switching to the AB configuration on 2005 January 14 and the source was detected during 10 epochs. The flux calibrators 3C~286, 3C~147 and/or 3C~48 were used depending on time and wavelength. Phase referencing was applied using the calibrators 1146+539 (JVAS~J1146+5356), 11534+49311 (JVAS~J1153+4931) or  11270+45161 (JVAS~J1126+4516), depending on wavelength. The data were reduced using standard flagging, calibration and imaging routines within the National Radio Observatory's Astronomical Image Processing System ( {\sc aips}). The flux densities of the primary calibrators were obtained using the formulae of Baars et al. (1977) but with the revised coefficients of Rick Perley, as is the default option in the {\sc setjy} routine. The flux densities of XTE~J1118+480 were calculated by fitting a Gaussian to the position of the source; each detection showed the source as an unresolved point source.

Additional radio observations at 15 GHz were obtained at the Ryle Telescope of the Mullard Radio Astronomy Observatory, Cambridge, UK. Further details of the observing technique can be found in Pooley \& Fender (1997). All VLA and Ryle flux densities are listed in Table~\ref{tab:radio}

\section{Results - Lightcurves}

The resultant lightcurves for the 2005 outburst of XTE~J1118+480 are shown in Fig.~\ref{fig:lightcurves}. There is good temporal coverage in all wavebands and there are many epochs with observations obtained simultaneously in more than one spectral region.

The top two panels of Fig.~\ref{fig:lightcurves} show the hard X-ray temporal behaviour as observed with BAT and HEXTE. The onset of the outburst was detected by the BAT instrument, showing a steep rise in the number of counts from MJD 53380 to a maximum around MJD 53384. HEXTE began pointed observations just prior to the hard X-ray peak (MJD 53384--5). They then decayed with at least one ``shoulder'' (e.g. $\sim$ MJD 53392) superimposed on the otherwise exponential decay profile (there are possibly two features on the HEXTE lightcurve -- $\sim$ MJD 53388 and MJD 53392). The final data-point of note is at $\sim$MJD 53410, an apparent reflare with possible related features at lower frequencies (see below). 

The soft X-rays at 2--12 keV (ASM: crosses) and 3--20 keV (PCA: solid circles), plotted in the third panel, show similar behaviour. They rise to a single peak, before decaying smoothly with a similar ``shoulder''-like feature superimposed on the PCA decay that coincides loosely with that of the BAT and HEXTE. The apparent rebrightening in the ASM data ($\sim$MJD 53412) and also in one of the PCU detectors on the PCA (PCU\,0 only, which we do not include in Fig.~\ref{fig:lightcurves}) appears to be a background event, despite the quasi-simultaneity with the features in HEXTE and the optical. 

\begin{figure}
\begin{center}
\leavevmode
\includegraphics[width=8cm]{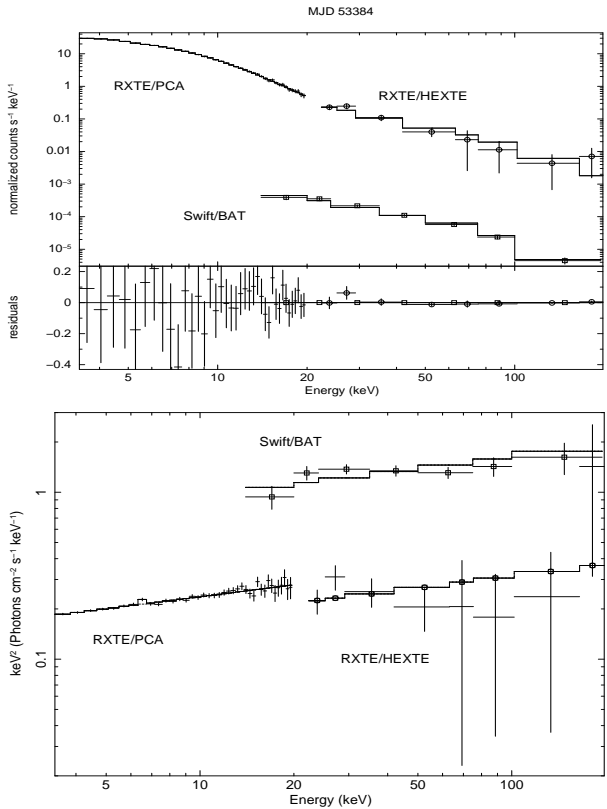} 
\caption{The combined {\sl RXTE}/PCA, {\sl RXTE}/HEXTE and {\sl Swift}/BAT data from MJD 53384 (53384a in Table~\ref{tab:xray-spectra}) fitted simultaneously with a simple power-law model. The data have been rebinned to 5$\sigma$ for the plot only. Top panel: counts spectrum with residuals, the PCA, HEXTE and BAT points are indicated by crosses, circles and squares respectively. Bottom panel: unfolded energy spectrum. The offset in normalisation between the BAT and HEXTE data is surprising and indicates the need for a careful study of spectral fits with the BAT survey data in relation to other instruments. Such study is beyond the scope of this paper but the discrepancy does not affect the results presented here.}
\label{fig:spectrum}
\end{center}
\end{figure}

\begin{table}
\begin{center}
\caption{Results of fitting a simple power-law spectral model to the combined {\sl RXTE}/PCA, {\sl RXTE}/HEXTE and {\sl Swift}/BAT data simultaneously. Where there is more than one {\sl RXTE} observation per day, we have used the single BAT spectrum with each {\sl RXTE} spectrum. We adopt $N_h=1.3\times 10^{20}$ and include a Gaussian of width 0.01 keV at 6.6 keV (following Maitra et al. 2009). The columns list values for the photon index and normalisation of the power law, the reduced $\chi^2$ and number of degrees of freedom. Errors are to within 90\% confidence. }
\label{tab:xray-spectra}
\begin{tabular}{lcccc}
\hline
\hline
Epoch  & PL Index & PL Normalisation &$\chi^2_{\nu}$&dof\\
\hline
53384a&$1.76\pm0.02$&$0.137\pm0.003$&0.75&73 \\
53388a&$1.77\pm0.01$&$0.101\pm0.002$&0.86&196\\
53388b&$1.76\pm0.01$&$0.095\pm0.002$&1.09&212\\
53390a&$1.79\pm0.02$&$0.086\pm0.003$&1.13&172\\
53390b&$1.78\pm0.01$&$0.094\pm0.002$&0.98&185\\
53390c&$1.78\pm0.02$&$0.089\pm0.002$&0.92&118\\
53392a&$1.79\pm0.03$&$0.069\pm0.004$&0.90&125\\
53392b&$1.79\pm0.02$&$0.071\pm0.002$&0.70&119\\
53392c&$1.84\pm0.02$&$0.066\pm0.003$&0.97&124\\
53394a&$1.83\pm0.01$&$0.054\pm0.001$&0.89&212\\
53394b&$1.83\pm0.02$&$0.044\pm0.001$&1.07&176\\
53394c&$1.83\pm0.03$&$0.049\pm0.002$&0.94&134\\
53394d&$1.84\pm0.03$&$0.050\pm0.002$&0.82&147\\
53396a&$1.84\pm0.02$&$0.039\pm0.001$&0.98&176\\
53396b&$1.83\pm0.04$&$0.033\pm0.002$&0.93&118\\
53398a&$1.91\pm0.09$&$0.025\pm0.004$&0.85&80 \\
53398b&$1.83\pm0.04$&$0.023\pm0.002$&1.00&163\\
\hline
\end{tabular}
\end{center}
\end{table}

\begin{figure}
\begin{center}
\leavevmode
\includegraphics[width=8cm]{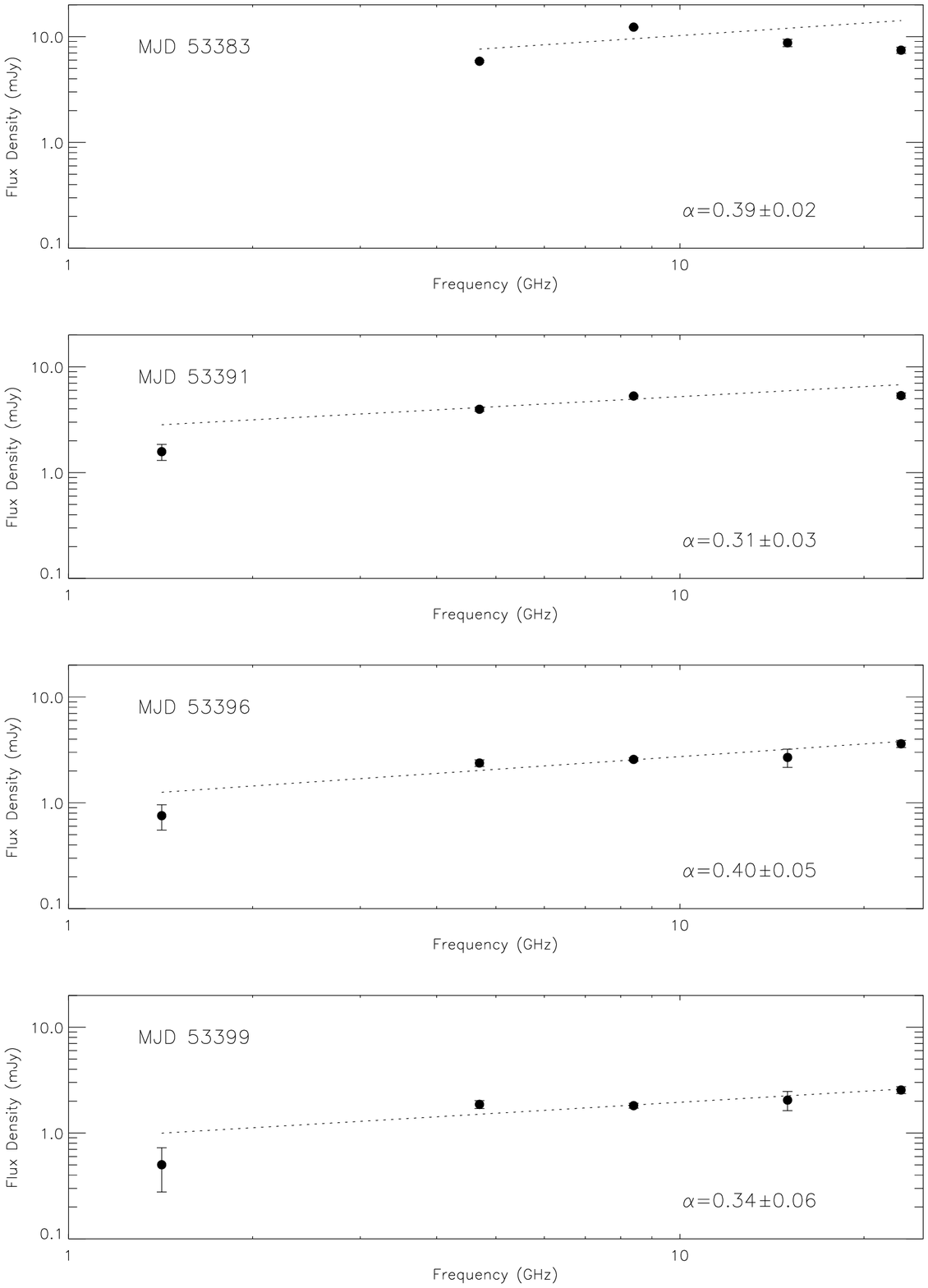}
\vspace*{0cm}
\caption{Radio spectra for those epochs with four or more frequencies. An optically thin turnover at higher frequencies can be seen in the top plot and may be ``contaminating'' the others. Best-fit straight lines of gradient $\alpha$ are plotted on each panel.}
\label{fig:alpha}
\end{center}
\end{figure}

The fourth panel shows the optical and infrared data and there is a very different lightcurve morphology; the decay is much slower than that of the X-rays. A second, smaller peak is observed in the $V$-band after the source had been decaying for about a month, although this occurs a few days after the reflare of the hard X-rays. We note that the BAT and ASM points before and after the outburst represent non-detections rather than detected quiescence and, as such, cannot be used to claim that the optical rise preceded that of the X-rays.

Finally the bottom panel shows the radio data. No points were obtained during the rise and so the radio must have peaked before or simultaneously with our first observation on MJD 53383. The radio peak therefore apparently precedes or coincides with the X-ray and optical peaks. The uncharacteristic broken decay of the better-sampled frequencies (4.7 and 8.4 GHz) hints at an additional ejection event, presumably related to the features seen in the X-ray lightcurves.  The radio source then decayed to below $3\sigma$ between MJD 53407 and 53414. 

\section{Results - Spectra}

Detailed broadband modelling of data from MJD 53393 can be found in Maitra et al. (2009). For this study we are interested only in the variability of the overall spectra; therefore we fit the soft/hard X-ray spectrum with a simple power-law, using the model {\sc powerlaw} within {\sc xspec V11.3.2}. This simple model was successful at fitting the hard X-ray spectra in the range 14--195 keV, yielding values of $\Gamma$ in the range 1.5--2.1. Similarly the model was successful at fitting the {\sl RXTE}/PCA spectra in the range 3--20 keV with $1.7<\Gamma<1.9$; we also included a Gaussian of width 0.01 keV and energy 6.6 keV, representing emission due to Fe K$\alpha$, which both improves the fit and provides consistency with Maitra et al. (2009). The resultant power-law fits are consistent with the X-ray source having remained in the low/hard state for the duration of the outburst. Having fit the hard and soft X-rays separately, we then fit {\sl Swift}/BAT, {\sl RXTE}/HEXTE and {\sl RXTE}/PCA simultaneously.  The resultant fit parameters are listed in Table~\ref{tab:xray-spectra} and show that this simple power-law model provides a good fit to the data. A sample spectrum is plotted in Fig.~\ref{fig:spectrum}. We see no evidence for an exponential cut-off to the hard X-ray spectrum up to 200 keV using the phenomenological cut-off power-law model, {\sc cutoffpl}.  Detailed broadband (radio to X-ray) spectral fits to quasi-simultaneous data at multiple epochs during the outburst decay will be presented in a forthcoming paper (Maitra et al. 2010, in prep.). 

The optical and infrared spectra have been analysed in depth by Zurita et al. (2006) and Hynes et al. (2006) and so we do not repeat their work here.

Radio spectral indices, $\alpha$ (defined by $S_{\nu}\propto\nu^{\alpha}$, where $S_{\nu}$ is the flux density at frequency $\nu$ ) have been calculated for each epoch for which there were two or more observing frequencies. They are listed in Table~\ref{tab:alpha} and the spectra plotted in Fig.~\ref{fig:alpha}. In each case, $\alpha$ is consistent with the inverted spectrum we would expect in the low/hard state. However, the high values of $\chi^2$ reflect the poor fits of the data to a power-law, particularly for MJD 53383, as can be seen in Fig.~\ref{fig:alpha}. It appears that although the spectra are predominantly partially self-absorbed, as indicated by the positive values of $\alpha$, there is also optically thin emission present at higher frequencies for some epochs. The observed radio spectrum is the composite of optically thin ($\alpha < 0$ ) radio emission from ejections earlier in the outburst that blend with the core flat-spectrum ($\alpha$ =0) radio emission (see e.g. Jonker et al. 2009). Fig.~\ref{fig:alpha} and Table~\ref{tab:alpha} suggest tentatively that the contribution of optically thin emission may be asscociated with the initial outburst and the radio ``shoulder'' at MJD 53391, when the power-law fits are worst, and decrease as the radio source decays. Again, however, the poor fits to the spectra require us to be cautious in drawing such a conclusion. 

\begin{table}
\begin{center}
\caption{Spectral indices for VLA data. The spectral index, $\alpha$, is defined in terms of $S_{\nu}\propto\nu^{\alpha}$, where $S_{\nu}$ is the flux density at frequency $\nu$. Epochs with two or more frequencies have been included. We note that the values of $\alpha$ may be misleading as the total radio emission may incude residual optically thin emission, from the onset of the outburst, at higher frequencies at some epochs. The relatively poor fits to a power-law are reflected in the high values of $\chi^2$ shown in the fourth column.}
\label{tab:alpha}
\begin{tabular}{lcccc}
\hline
\hline
MJD&$\alpha$&$\alpha_{\mbox{err}}$&$\chi^2$&DOF\\
\hline
53383.3& 0.39&    0.02&      743.0 &          2    \\  
53384.7& 0.42&    0.07&      1.9 &          0     \\
53386.2& 0.43&    0.05&      3.4 &          1     \\
53391.6& 0.31&    0.03&      54.1 &          2     \\
53394.3& 0.41&    0.05&     0.3 &          1     \\
53396.5& 0.40&    0.05&      11.9 &          3     \\
53399.3& 0.34&    0.06&      10.3 &          3     \\
53404.2& 0.37&    0.18&      4.8 &          1     \\
53407.5& 0.31&    0.94&     0.7 &          0     \\
\hline
\end{tabular}
\end{center}
\end{table}

\section{Results - flux:flux correlations}
\begin{figure}
\begin{center}
\leavevmode
\includegraphics[width=6cm,angle=90]{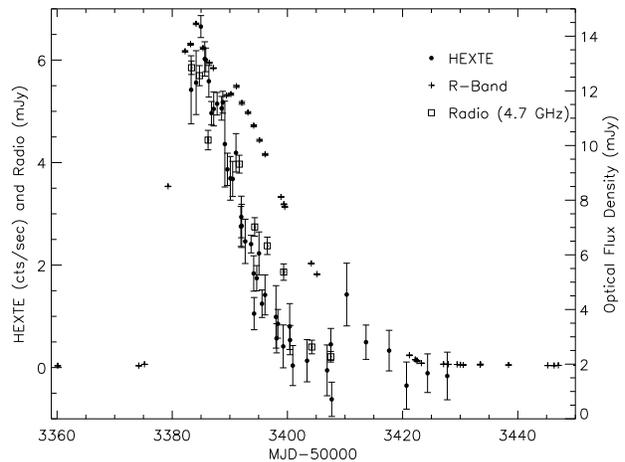}
\vspace*{0cm}
\caption{Superimposed HEXTE, radio and $R$-band lightcurves; the $R$-band points have been converted into dereddened flux density for more accurate comparison. There is significant deviation between the $R$-band and the other lightcurves.} 
\label{fig:lc}
\end{center}
\end{figure}

\begin{figure}
\begin{center}
\leavevmode
\includegraphics[width=9cm]{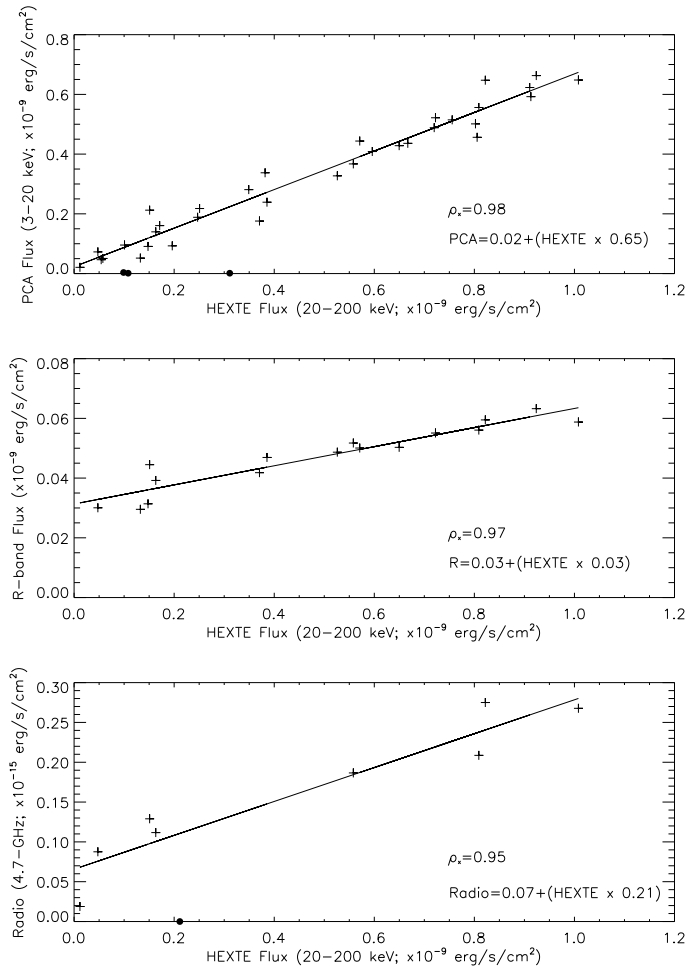}
\vspace*{0cm}
\caption{Flux:flux plots for HEXTE:PCA, HEXTE:$R$-band and HEXTE:radio(4.7 GHz). Spearman rank correlation components are shown on each panel; $\rho\ge 0.95$ to within 99\% confidence in each case. The best fit straight lines are plotted over the data, accompanied by their respective equations. The three filled circles are additional (unfitted) points, the faint X-ray fluxes of which were approximated using {\sc WebPimms}. Two of these three points (filled circles) lie off the correlations, reflecting the HEXTE reflare.}
\label{fig:correlation}
\end{center}
\end{figure}

In Section 3 we commented on the degree of flux correlation between the various wavebands and investigate it further here. We use the HEXTE data as a benchmark lightcurve since it has well-sampled data and is least likely to be contaminated with disc emission. Using $E(B-V)=0.21$ (Hynes et al. 2000), the extinction law of Cardelli, Clayton \& Mathis (1989) and the flux conversions of Bessell, Castelli \& Plez (1998), we converted the $R$-band brightness values to extinction-corrected flux density and overplotted the HEXTE, $R$-band and 4.7-GHz lightcurves (Fig.~\ref{fig:lc}). It is immediately apparent that the X-ray and radio lightcurves have similar slopes, while the $R$-band deviates during its decay. In order to quantify this correlation, we used the power-law spectral fits to determine the HEXTE and PCA flux and plot flux:flux plots for HEXTE:PCA, HEXTE:$R$-band and HEXTE:radio(4.7-GHz) (Fig.~\ref{fig:correlation}). Data-pairs were as simultaneous as possible and always to within one day. The resultant Spearman correlation coefficients, $\rho$, are $\ge 0.95$ to within 99\% confidence. Such a high correlation is surprising for the $R$-band and may reflect a change in the spectrum (to be investigated further in a future work) -- from the lightcurves we suspect that the correlation would worsen as time progresses. The filled circles indicate extra points (not included in the fit) for which HEXTE and PCA spectra could not be obtained due to the low count-rates; approximate fluxes were derived using {\sc WebPimms} and assumed photon-index of 1.8. The two points which fall off the correlation reflect the HEXTE reflare.

\section{Discussion}

The lightcurves of the 2005 outburst of XTE J1118+48 are notable for their FRED-like (fast rise, exponential decay) morphology. They show a high degree of correlation between the radio and X-ray emission, suggesting a close relation between the emission processes and/or emitting regions (see Zdziarski \& Gierlinski (2004) for discussion of how the correlation does not {\em require} the X-ray and radio emission to be produced via the {\em same} process). The optical lightcurve decays more gradually than those of the other bands. There also appears to be an optically thin component to the radio emission, which possibly decays in favour of the self-absorbed jet as the outburst proceeds. 

The radio emission can be attributed to a jet on account of its self-absorbed spectrum (Fender 2006 and references therein). If the jet were to dominate the optical and infrared emission then we would expect to see a greater correlation between their respective lightcurves. Indeed, the lightcurves shown here could lead us to assume that the optical and infared are dominated by disc emission were it not for the additional synchrotron components found by Zurita et al. (2006), Hynes et al. (2006) and Maitra et al. (2009). Nonetheless, despite the known additional synchrotron component, the excess of optical/infrared emission following the X-ray/radio decay shows that the disc emission is still likely to be significant at these wavelengths.

There has been recent debate regarding whether there can be a disc component to the X-ray spectrum when a source is in the low/hard state (e.g. Reis et al. 2009; Rykoff et al. 2007; Miller et al. 2006; Chiang et al. 2009). Gierli\'nski, Done \& Page (2008, 2009) have explained the softening of the spectrum in terms of irradiation of the inner disc and Done \& D\'{i}az Trigo (2009) demonstrated that the iron line is an artefact of pile-up, thereby resolving the controversy over whether or not the disc is truncated. This is also discussed in Maitra et al. (2009), whose modelling of one epoch during the decay of XTE J1118+480 in 2005 allows for (although does not require) a small inner disc radius and, consequently, a source of soft X-rays with which to irradiate the outer disc. For a typical low/hard state outburst we would expect the power-law component of the X-ray emission to dominate at both high and low energies and for there to be little difference between the respective lightcurves. That indeed appears to be the case in the high and low energy X-ray lightcurves presented here and there is no suggestion that there are contributions from more than one component (contrasting with the lightcurves of, for example, GRO~J1655$-$40; Brocksopp et al. 2006).

Once the X-rays and radio have started to decay, their behaviour is closely linked. They all appear to undergo some additional event around MJD 53385--53390, manifested as a ``shoulder'' superimposed on the decay. The profile of the radio lightcurve is more comparable with that of sources which made transitions to softer spectral states, such as XTE~J1720$-$318 or A0620$-$00 (Brocksopp et al., 2005; Kuulkers et al. 1999 respectively) than we would expect for a source in the low/hard state. A second, optically-thin, jet ejection event seems a likely cause of this radio ``shoulder'', particularly as the radio spectrum at this time is particularly poorly fit by a power-law. Confirmation of this would require detection of a simultaneous reduction in the spectral index, needing higher sensitivity and sampling in the radio data. Multiple ejections are commonly associated with transient events (Brocksopp et al. 2002) but more usually with the optically thin events of sources which soften and enter the very high state (Fender et al. 2004). Such jet ejections may be unexpected during the low/hard state but not unprecedented (e.g. GS~1354$-$64 Brocksopp et al. 2001). It may be that the ``canonical'' stable, compact flat-spectrum jet of the low/hard state (e.g. Cyg~X-1, GX~339$-$4 or the 2000 outburst of XTE~J1118+480; Markoff et al. 2005 and references therein) may turn out to be the exception rather than the norm. Alternatively, this ``ejection'' may be more like the flare seen in V404 Cyg during quiescence, thought to be some sort of re-energising of the electrons within the jet rather than a new ejection event (Miller-Jones et al. 2008), albeit on a much longer timescale.

Given the link between the power-law X-ray emission and the jet it might seem surprising that the reflare at $\sim$MJD 53415 was detected at optical and hard X-ray wavelengths but not the radio. We note that the X-ray reflare preceded the optical reflare and occurred in a gap between radio observations and so we could have missed a similar event in the radio due to the sparse sampling. Multiwavelength daily monitoring of these events is required at high sensitivity to determine their true nature.

Finally, we compare these results with the X-ray, optical and radio lightcurves of the 2000 outburst of XTE J1118+480, detailed analyses of which were presented by Chaty et al. (2003) and Brocksopp et al. (2004). The lightcurves of that event are notable for their highly correlated behaviour at all frequencies, with the long, plateau-like second peak seen simultaneouly at hard X-ray, soft X-ray, optical and radio wavelengths. Spectral fits showed that the synchrotron jet was found to extend to high frequencies, possibly the hard X-rays, and dominate the contribution from any disk component (Markoff et al. 2001). In contrast, the 2005 event shows a much more ``canonical'' FRED lightcurve morphology, often seen in soft X-ray transient events. The optical emission shows a much slower decay than the X-ray and radio, suggesting that it is dominated by an alternative component, most likely an accretion disc, although there also appears to be a synchrotron component (as discussed above). The flat spectrum of the synchrotron emission is contaminated by optically thin radio emission. Different phenomena are clearly dominating the 2000 and 2005 events, despite both being governed by the properties of the low/hard state. Obviously these phenomena cannot be properties which would remain unchanged over the intervening 5 years, such as orbital parameters, mass of the components or the spin of the black hole. We may need to consider properties intrinsic to the accretion disc or jets instead.

\section{Conclusions}
XTE~J1118+480 has now been observed during two outbursts. We have obtained data for the 2005 event, covering all wavebands at multiple epochs. We use analysis of the lightcurves to disentangle jet and disc emission with a view to full spectral modelling in a future work. The results point towards a very different nature of XTE~J1118+480 in 2005 compared with 2000, despite both events remaining in the low/hard state and showing correlated X-ray and radio behaviour. In 2000 the lightcurves at X-ray, optical and radio wavelengths were characterised by a long, plateau-like phase, which could be attributed to a dominant and long-lasting stable jet, with a flat spectrum extending at least through the lower frequencies. In contrast, the 2005 outburst behaved more like a canonical soft X-ray transient, with short-lived jet ejection events, an optically thin contribution to the synchrotron spectrum, a more dominant disc component and a ``fast rise, exponential decay'' lightcurve morphology. These results add to the discussion that the low/hard state covers a wider range of properties than typically assumed. Furthermore, they show that the nature of an outburst in the low/hard state must be governed by at least one property that can vary on a timescale of just a few years.

\section*{Acknowledgments}

We thank Sandy Leggett for help with the UKIRT data reduction, J\"orn Wilms for advice regarding the HEXTE analysis and the anonymous referee for useful suggestions which improved the paper. PGJ acknowledges support from a VIDI grant from the Netherlands Organisation for Scientific Research. HAK is supported by the Swift project. This research has made use of data obtained through the High Energy Astrophysics Science Archive Research Center Online Service, provided by the NASA/Goddard Space Flight Center. The Liverpool Telescope is operated on the island of La Palma by Liverpool John Moores University in the Spanish Observatorio del Roque de los Muchachos of the Instituto de Astrofisica de Canarias with financial support from the UK Science and Technology Facilities Council. The United Kingdom Infrared Telescope is operated by the Joint Astronomy Centre on behalf of the Science and Technology Facilities Council of the U.K. The VLA is a facility of the National Radio Astronomy Observatory, which is operated by Associated Universities Inc., under cooperative agreement with the National Science Foundation.

\end{document}